\documentclass[twocolumn,amsmath,amssymb]{revtex4-1}
\usepackage[dvipdfmx]{graphicx} 
\usepackage{dcolumn} 
\usepackage{bm} 
\usepackage{braket} 
\usepackage[usenames,dvipsnames]{color}
\usepackage[
colorlinks=true, citecolor=blue, urlcolor=blue, linkcolor=blue,
setpagesize=false, bookmarks=false
]{hyperref}

\begin{document}
\title{
Excitonic effects on high-harmonic generation in Mott insulators
}
\author{Mina Udono$^{1}$, Koudai Sugimoto$^{2}$, Tatsuya Kaneko$^{3}$, and Yukinori Ohta$^{1}$}
\affiliation{
$^1$Department of Physics, Chiba University, Chiba 263-8522, Japan\\
$^2$Department of Physics, Keio University, Yokohama, Kanagawa 223-8522, Japan\\
$^3$Computational Quantum Matter Research Team, RIKEN Center for Emergent Matter Science (CEMS), Wako, Saitama 351-0198, Japan
}
\date{\today}

\begin{abstract}
To study excitonic effects on high-harmonic generation (HHG) in Mott insulators, we investigate pumped nonequilibrium dynamics in the one-dimensional extended Hubbard model.
By employing time-dependent calculations based on the exact diagonalization and infinite time-evolving block decimation methods, we find the strong enhancement of the HHG intensity around the exciton energy.
The subcycle analysis in the sub-Mott-gap regime shows that the intensity region of the time-resolved spectrum around the exciton energy splits into two levels and oscillates following the driving electric field.
This excitonic dynamics is qualitatively different from the dynamics of free doublon and holon but favorably contributes to HHG in the Mott insulator.
\end{abstract}

\maketitle

\section{Introduction}
Recent advances in laser techniques, which can generate high-intensity and ultra-short optical pulses, have enabled the observation of a variety of nonlinear optical responses~\cite{RevModPhys.81.163}.
Among them, high-harmonic generation (HHG) is important in terms of application, e.g., to attosecond light sources~\cite{RevModPhys.72.545}.
Its physical process also attracts interest because HHG can reflect underlying electronic properties.
While HHG in atoms and molecules have been well-established~\cite{PhysRevLett.71.1994,PhysRevA.49.2117}, HHG in bulk solids was reported during the past decade~\cite{Ghimire2011,Schubert2014,Hohenleutner2015,Vampa2015,Ndabashimiye2016,Liu2017} and  these experimental achievements stimulate many theoretical studies~\cite{PhysRevLett.113.213901,PhysRevB.91.064302,PhysRevLett.116.016601,PhysRevA.98.023415,PhysRevA.94.063403,PhysRevA.95.043416,PhysRevA.96.053418,PhysRevLett.113.073901}.
Since HHG in bulk materials can capture dynamical properties of Bloch electrons, the techniques for detecting the electronic band structures~\cite{,Luu2015,PhysRevLett.115.193603,PhysRevLett.124.157403}, Berry curvature~\cite{Luu2018}, and transition dipole moment~\cite{PhysRevB.103.L161406} have been proposed.
While HHG in semiconductors and semimetals, for which the single-particle band picture is valid, have been investigated intensively, many-body effects on HHG in strongly correlated systems attracts attention recently~\cite{Silva2018,PhysRevLett.121.057405,PhysRevB.98.075102,PhysRevB.101.195139,PhysRevLett.124.157404,PhysRevB.103.035110,PhysRevResearch.3.023250,PhysRevB.103.035110,PhysRevB.104.245103,orthodoxou2021,PhysRevLett.128.047401,PhysRevLett.128.127401,arxiv.2203.01029}.
The previous studies point out that motions of quasiparticles associated with correlation effects, e.g., doublon (doubly occupied site) and holon (empty site) in Mott insulators (MIs), play a key role in HHG~\cite{PhysRevLett.124.157404,PhysRevB.103.035110}.

\begin{figure}[b]
\begin{center}
\includegraphics[width=\columnwidth]{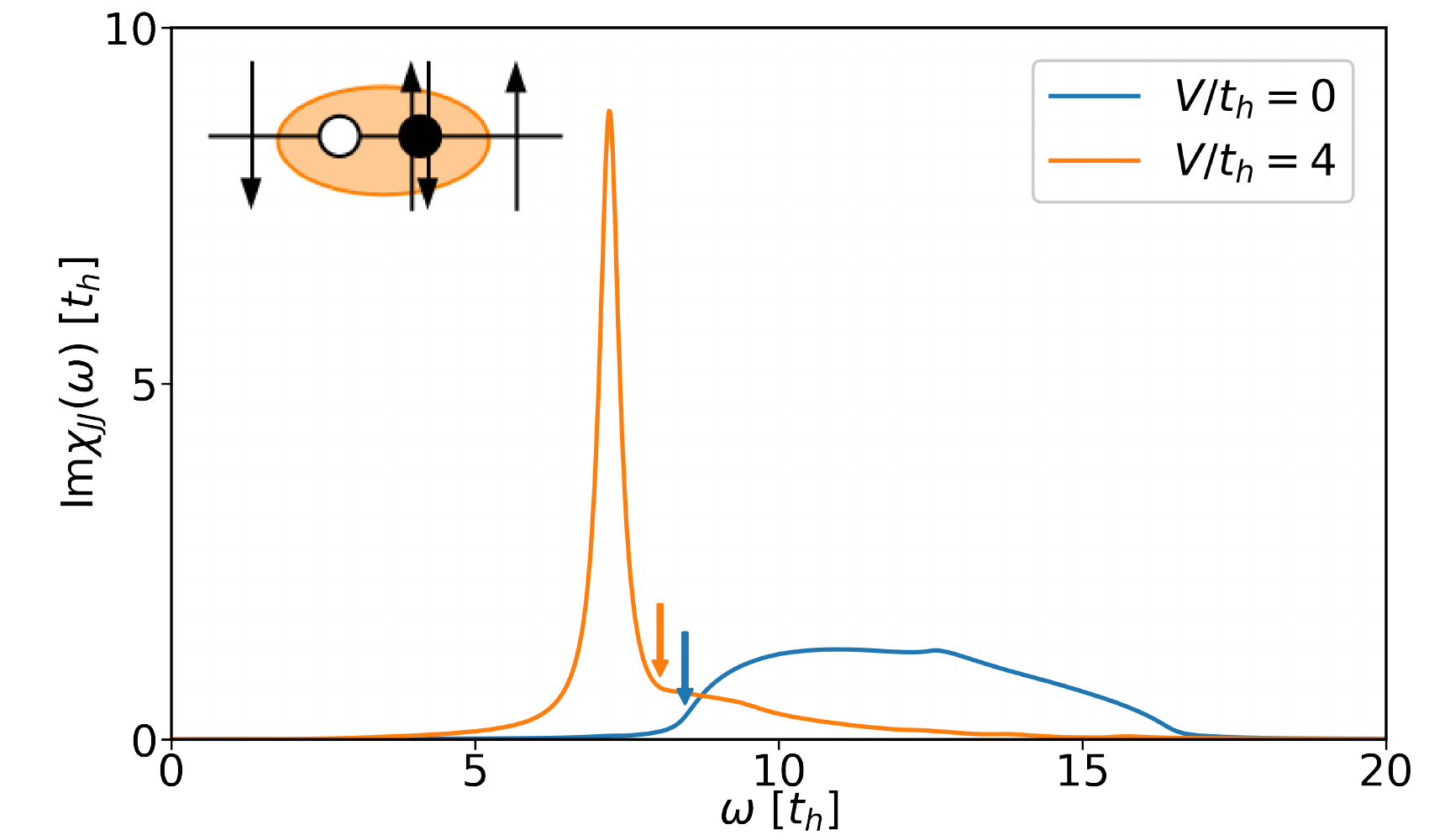}
\caption{
Imaginary parts of the linear optical response function calculated by the TEBD method with the infinite-boundary condition for $U/t_h=12$ with $V/t_h=0$ (blue) and $V/t_h = 4$ (orange),  where a broadening factor $\eta/t_h = 0.2$ is used.
The time evolutions of the window states are
carried out up to $t = 50 /t_h$.
The vertical blue and orange arrows indicate the Mott gap for $V/t_h=0$ and $V/t_h=4$, respectively.
Inset: Schematic picture of the exciton (doublon-holon pair) in the MI.
}
\label{fig:LinearRF}
\end{center}
\end{figure}

When nonlocal interactions are crucial in a correlated system, quasiparticles compose a bound state in its optical excitation process.
In the case of the MI, the doublon and holon make the composite particle, exciton, due to the inter-site Coulomb interaction $V$ (see also Fig.~\ref{fig:LinearRF})~\cite{PhysRevB.54.R17269,PhysRevB.55.15368,PhysRevB.56.15025,PhysRevB.64.125119,PhysRevB.67.075106}.
The optical experiments for the one-dimensional MIs, which exhibit the strong third-harmonic responses,  have suggested the importance of excitonic effects~\cite{Kishida2000,PhysRevLett.85.2204,PhysRevB.70.085101,PhysRevLett.95.087401}.
While we expect that the exciton in the MI contributes to HHG, its mechanism should be different from HHG in the simple MI (at $V=0$) because the motions of the doublon and holon are strongly restricted by the doublon-holon interaction.
However, excitonic effects on HHG in MIs have not so far been studied theoretically.

To address this issue, we consider the exciton in the MI described by the one-dimensional extended Hubbard model and investigate pumped nonequilibrium HHG dynamics by employing the time-dependent calculations based on the exact diagonalization (ED) and infinite time-evolving block decimation (iTEBD) methods.
We demonstrate that the HHG intensity in the MI is strongly enhanced around the exciton energy in the sub-Mott-gap regime.
In addition, our subcycle analysis shows that the intensity region of the time-resolved spectrum around the exciton energy splits into two levels and oscillates following the driving electric field.
While this excitonic dynamics is qualitatively different from the dynamics of free doublon and holon,  the exciton in the MI favorably contributes to HHG.

\section{Model and Methods}
To study excitonic effects, we introduce the one-dimensional extended Hubbard model described by
\begin{align}
\hat{H}=&-t_{h}\sum_{j,\sigma}(\hat{c}^{\dagger}_{j,\sigma}\hat{c}_{j+1,\sigma}+{\rm H.c.})
\notag \\
 &+U\sum_{j}\hat{n}_{j,\uparrow}\hat{n}_{j,\downarrow} + V\sum_{j}\hat{n}_{j}\hat{n}_{j+1},
\end{align}
where $\hat{c}^{\dag}_{j,\sigma}$ ($\hat{c}_{j,\sigma}$) is the creation (annihilation) operator for a fermion at site $j$ with spin $\sigma$ ($=\uparrow,\downarrow$), and $\hat{n}_{j,\sigma} = \hat{c}^{\dag}_{j,\sigma} \hat{c}_{j,\sigma}$ ($\hat{n}_{j} = \hat{n}_{j,\uparrow} + \hat{n}_{j,\downarrow}$).
$t_h$ is the hopping amplitude between the nearest-neighbor sites while $U$ and $V$ are the on-site and nearest-neighbor repulsive interactions, respectively.
We consider the case at half-filling, where the total number of particles $N$ is equal to the system size $L$.
In this case, the ground state is the antiferromagnetic MI state (spin density wave) at $U > 2V$ while the charge density wave is stabilized at $U < 2V$ in the large-$U$ limit~\cite{PhysRevB.61.16377,PhysRevLett.88.056402,PhysRevLett.99.216403}.
Furthermore, when $V > 2t_h$ in the MI phase, the excitonic peak emerges in the optical excitation spectrum~\cite{PhysRevB.54.R17269,PhysRevB.55.15368,PhysRevB.56.15025,PhysRevB.64.125119,PhysRevB.67.075106}.

To discuss a pumped dynamics for HHG, we introduce a time-dependent external field via the Peierls phase by replacing $t_h \hat{c}^{\dagger}_{j,\sigma}\hat{c}_{j+1,\sigma} \rightarrow t_h e^{-iqA(t)}\hat{c}^{\dagger}_{j,\sigma}\hat{c}_{j+1,\sigma}$, where $A(t)$ is the vector potential and $q$ is the charge of a particle.
The electric field $E(t)$ is equal to $-\partial_t A(t)$~\cite{cha}.
In this paper, we use $A(t)=E_0/\omega_{\mathrm{p}}e^{-(t-t_0)^2/2\sigma_{\mathrm{p}}^2}{\rm sin}[\omega_{\mathrm{p}}(t-t_0)]$ assuming the amplitude of the electric field $E_0$ with the frequency $\omega_{\rm p}$ and the pulse width $\sigma_{\rm p}$ centered at time $t_0$~\cite{PhysRevB.103.035110}.
The pump frequency $\omega_{\rm p}$ is the fundamental frequency of higher-harmonics characterized by $\omega = n \omega_{\rm p}$ ($n$: integer).
To evaluate the HHG spectrum, we calculate the time-dependent current $J(t)$, whose operator is given by $\hat{J}(t)=i q t_h\sum_{j,\sigma}(e^{iqA(t)}\hat{c}^{\dagger}_{j+1,\sigma}\hat{c}_{j,\sigma}-e^{-iqA(t)}\hat{c}^{\dagger}_{j,\sigma}\hat{c}_{j+1,\sigma})$.
Unless otherwise noted, the quantity $J(t)$ denotes the current per site $\braket{\hat{J}(t)}/L$.
Then, performing the Fourier transformation of $J(t)$, we evaluate the HHG spectrum $I(\omega)=|\omega J(\omega)|^2$.
Here, we assume that the acceleration of charges leads to the emitted radiation.

In our simulations, the initial state is the ground MI state (at $U>2V$), and the state $\ket{\Psi(t)}$ under the external field is obtained by solving the time-dependent Sch\"odinger equation numerically.
In the finite system, we employ  the exact diagonalization (ED) method for the ground state and use the Krylov subspace technique for the time evolution $\ket{\Psi(t+\delta t)} \simeq e^{-i \hat{H}(t) \delta t}\ket{\Psi(t)}$ with a short time step $\delta t$~\cite{doi:10.1063/1.451548,MOHANKUMAR2006473,PhysRevB.96.235142}.
In the ED calculations, we use the $L=10$ site cluster with the periodic boundary condition.
We set the time step to be $\delta t = T_{\rm p} / m < 0.001/t_h $ (where $T_{\rm p}=2\pi/\omega_{\rm p}$ and $m$ is an integer) and the order of the  Krylov subspace for the time evaluation is $M=15$.

We also employ the iTEBD method~\cite{PhysRevLett.98.070201, Orus2008PRB} for the calculations in the infinite size system.
We obtain the ground state by the imaginary-time evolution, and for the real-time evolution of the pump dynamics, we use a fourth-order Trotter decomposition with the time step $\delta t = 0.005 / t_{\mathrm{h}}$.
We set the maximum bond dimension $\chi = 1000$, which the obtained results are sufficiently converged.
The linear response function $\chi_{JJ} (\omega) = ( i/L ) \int^{\infty}_0 \braket{\psi_0 | [\hat{J}_{\rm I}(t),\hat{J}_{\rm I}(0)] |\psi_0} e^{i\omega t - \eta t} dt$ of the ground state $\ket{\psi_0}$ (where $\hat{J}_{\rm I}(t)=e^{i\hat{H}t} \hat{J} e^{-i\hat{H}t}$ indicates the interaction picture) is calculated by the TEBD method with the infinite-boundary condition~\cite{Phien2012PRB, Phien2013PRB, 10.21468/SciPostPhys.10.3.077} in the uniform update scheme~\cite{Zauner2015JPCM, PhysRevResearch.4.L012012, arxiv.2204.09085}, where we use the window size $L_{\mathrm{w}} = 128$.
The damping $e^{-\eta t}$~\cite{damp}  
is included in the response function because the numerical simulation is restricted to finite time, and this leads to a Lorentzian broadening in the frequency space. 
In our calculations, we set $t_h$ ($t_h^{-1}$) as a unit of energy (time) and use $q=-1$.
Since the actual numerical calculations for the HHG spectrum $I(\omega)$ are performed in a finite time range $[0,t_{\rm max}]$, we introduce a Gaussian window function $F_{\rm gauss}(t)=\frac{1}{\sqrt{2\pi}\sigma_{\rm w}}\exp\left[-\frac{(t-t_{0})^{2}}{2\sigma_{\rm w}^2}\right]$ in the Fourier transformation of $J(t)$, i.e., $J(\omega)= \int_{0}^{t_{\rm max}}J(t)F_{\rm gauss}(t)e^{i\omega t}dt$~\cite{PhysRevB.103.035110}. 
We set $\sigma_{\rm w}$, $t_0$, and $t_{\rm max}(=2t_0)$ to be enough larger than the pulse width $\sigma_{\rm p}$ and the parameters used in the Fourier integral for the HHG spectrum $I(\omega)$ does not change our main results qualitatively.

\begin{figure}[t]
\begin{center}
\includegraphics[width=\columnwidth]{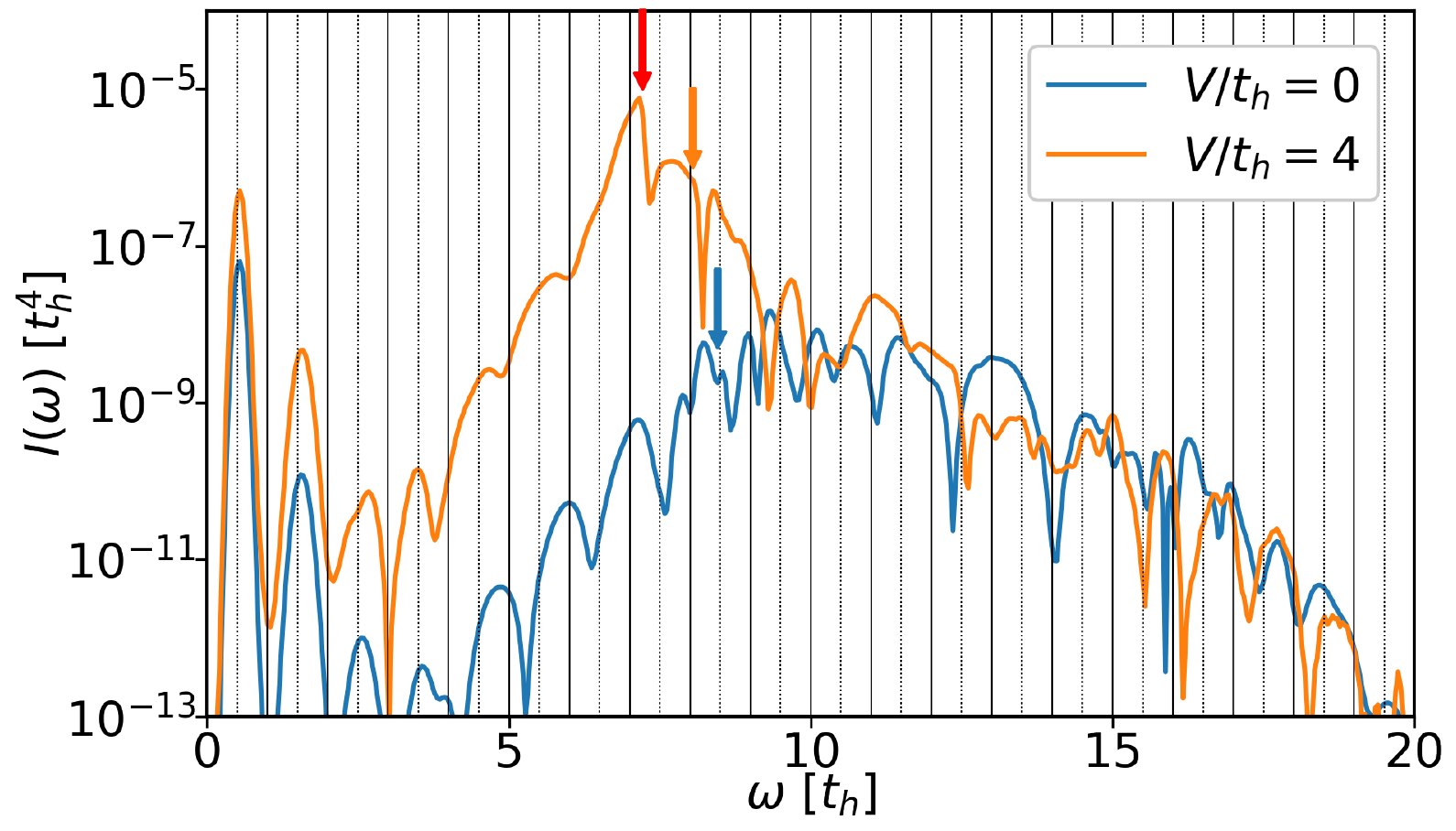}
\caption{
HHG spectra evaluated by the iTEBD method for $U/t_h=12$ with $V/t_h=0$ and $V/t_h=4$.
The vertical black solid (dotted) lines indicate even (odd) harmonics of $\omega_{\rm p}$.
The blue and orange arrows indicate the Mott gaps for $V/t_h=0$ and $V/t_h=4$, respectively, while the red arrow indicates the exciton energy for
$V/t_h=4$.
$\omega_{\rm p}/t_h=0.5 $, $E_0 /t_h = 1$, $\sigma_{\mathrm{p}}=T_{\mathrm{p}}$, and $t_0=5T_{\mathrm{p}}$ are used in the pump pulse.
The width of the Gaussian window function is $\sigma_{\rm w} = 1.8 T_{\rm p}$.
}
\label{fig:HHG_iTEBD}
\end{center}
\end{figure}

\section{Results}
Figure~\ref{fig:LinearRF} shows the imaginary parts of the linear response function $\chi_{JJ} (\omega)$ with $\eta/t_h = 0.2$.
The orange arrow in Fig.~\ref{fig:LinearRF} represents the Mott (charge) gap $\Delta_{\rm M} (L) = E_{\rm gs}(L+1) + E_{\rm gs}(L-1) - 2 E_{\rm gs}(L)$ in the thermodynamic limit $L \to \infty$~\cite{PhysRevB.67.075106,PhysRevB.68.045110}, where $E_{\rm gs}(N)$ is the ground-state energy for the $N$-particle system (see Ref.~\cite{Mgap} for details).
This Mott gap $\Delta_{\rm M}$ corresponds to the bottom of the doublon-holon continuum.
As shown in Fig.~\ref{fig:LinearRF}, while the optical spectrum for $V=0$ has the weight only above the Mott gap, the spectrum for $V/t_h=4$ exhibits the exciton peak below the Mott gap as in the previous studies~\cite{PhysRevB.64.125119,PhysRevB.67.075106}.
This result indicates that the exciton formed by $V$ considerably changes the optical properties of the MI.
We note that while the exciton peak is broadened by $\eta$, the frequency of the peak position is almost independent of small $\eta$ \cite{PhysRevB.67.075106}, and we check that $\eta/t_h=0.2$ is small enough.

To see the excitonic effect on HHG, we compare the HHG spectra $I(\omega)$ with and without the interaction $V$.
Figure~\ref{fig:HHG_iTEBD} shows the HHG spectra evaluated by iTEBD.
The red arrow is the exciton energy observed in the linear response function in Fig.~\ref{fig:LinearRF}.
The pump frequency $\omega_{\rm p} (=0.5t_h)$ is much smaller than the band gap as in the experiments for semiconductors~\cite{Ghimire2011}.
When $V=0$, while the intensities at the lower-order odd harmonics [i.e., $I(\omega \!=\! (2n+1)\omega_{\rm p})$] once decrease with $\omega$, the HHG response grows up with approaching the Mott gap $\Delta_{\rm M}$ (blue arrow in Fig.2) and $I(\omega)$ at $\omega > \Delta_{\rm M}$ exhibits the plateau structure, in which the HHG intensities hardly decay even when the harmonic order is increased~\cite{PhysRevLett.121.057405,PhysRevB.103.035110}.
This behavior is qualitatively consistent with the results in the previous studies~\cite{Silva2018,PhysRevB.103.035110}.
When $V/t_h=4$, in addition to the plateau-like structure above the Mott gap (orange arrow), the HHG intensity is enhanced around the exciton energy (red arrow).
The enhanced $I(\omega)$ in the sub-Mott-gap regime ($\omega < \Delta_{\rm M}$) implies that the exciton in the MI can be a good ingredient for HHG.
Note that, while all intensity peaks are expected to be centered at the odd harmonics in a system with inversion symmetry, we find deviations of peaks from the odd harmonics.
As discussed in Ref.~\cite{PhysRevB.103.035110}, this is probably because the system does not reach a time-periodic steady state for the reasons that the pulse is not long enough and that dephasing effects are missing in our simulation.

\begin{figure}[t]
\begin{center}
\includegraphics[width=\columnwidth]{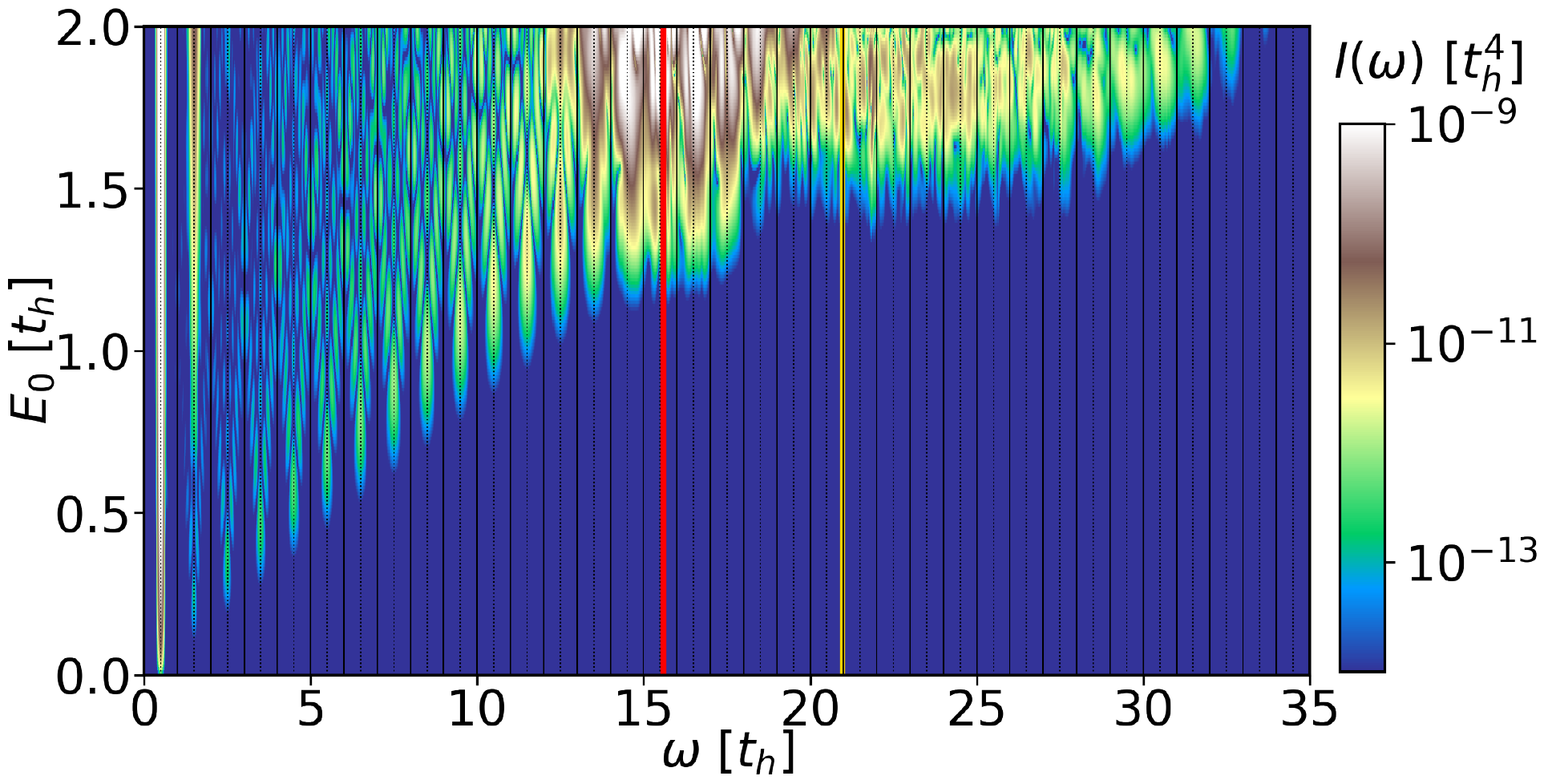}
\caption{
HHG spectrum in the plane of $\omega$ and $E_0$ evaluated by the ED method for $U/t_h=24$ and $V/t_h=8$. 
$\omega_{\mathrm{p}}/t_h=0.5$, $t_0=12T_{\mathrm{p}}$, $\sigma_{\mathrm{p}}=1.8T_{\mathrm{p}}$, and $\sigma_{\rm w}=3.6T_{\mathrm{p}}$ are used.
The vertical red and orange lines indicate the exciton energy and Mott gap, respectively.
}
\label{fig:HHG_ED}
\end{center}
\end{figure}

\begin{figure}[!t]
\begin{tabular}{cc}
	\begin{minipage}[t]{1\hsize}
	\centering
	\includegraphics[width=1\textwidth]{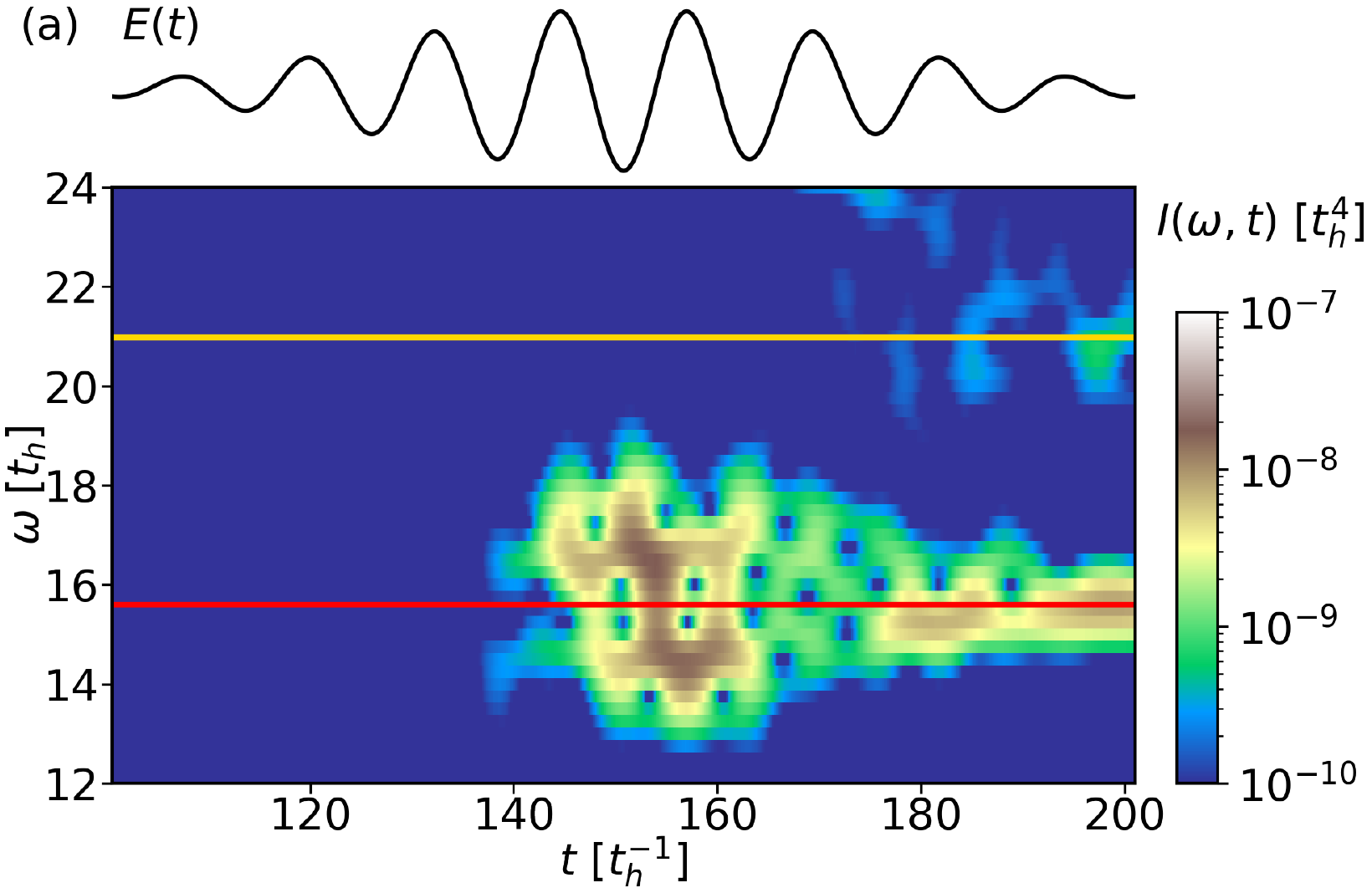}
	\end{minipage} \\
	\begin{minipage}[t]{1\hsize}
	\centering
	\includegraphics[width=1\textwidth]{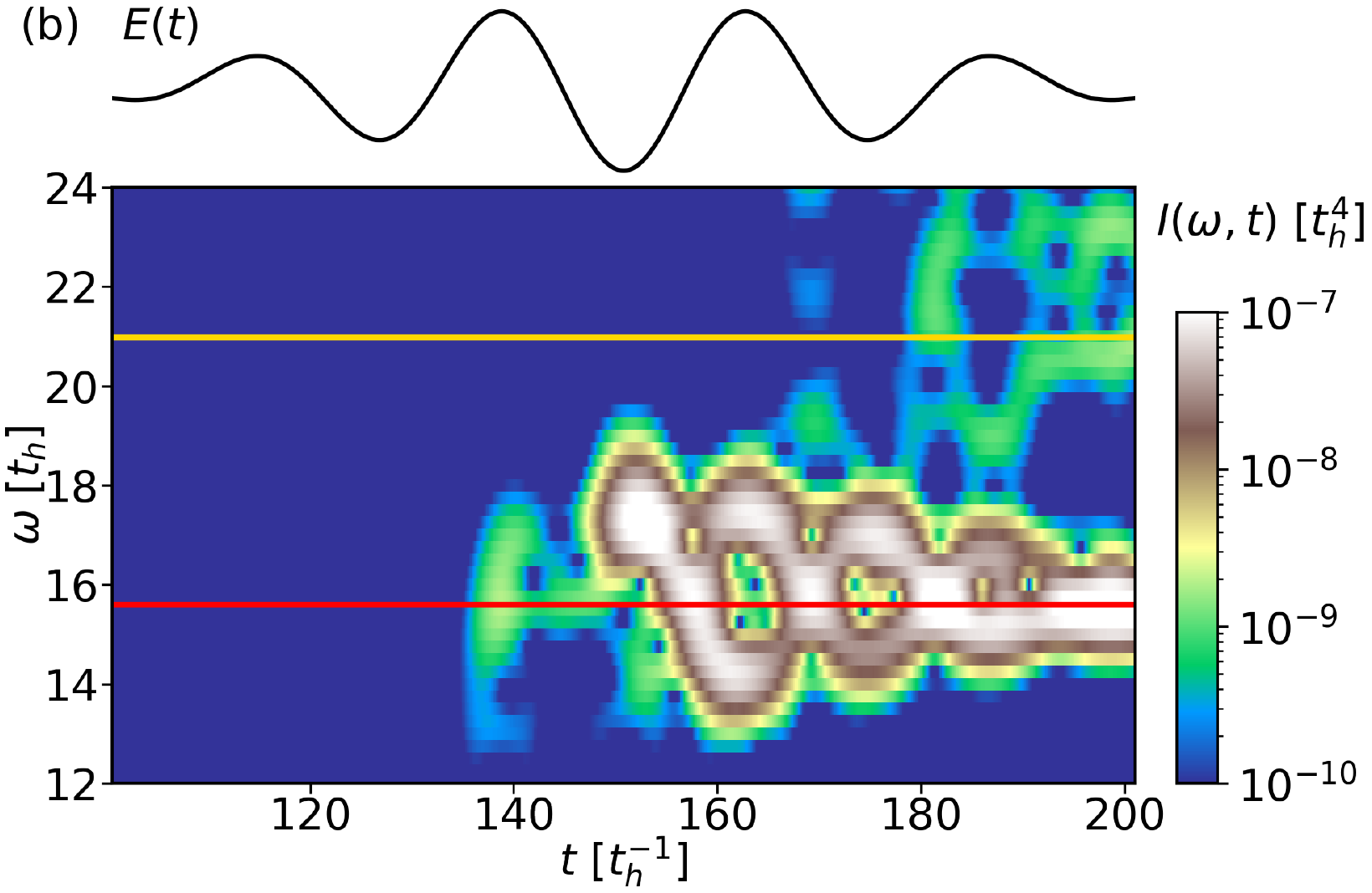}
	\end{minipage}
\end{tabular}
\caption{
Subcycle analysis performed by the ED method for $U/t_h=24$ and $V/t_h=8$. 
Time-resolved spectrum $I(\omega,t)$ is calculated at (a) $\omega_{\mathrm{p}}/t_h=0.5$ and (b) $\omega_{\mathrm{p}}/t_h=0.25$ with $E_0/t_h=1.75$.
$t_0=12T_{\mathrm{p}}$, $\sigma_{\mathrm{p}}=1.8T_{\mathrm{p}}$, and $\sigma_{\rm w}'=0.16T_{\mathrm{p}}$ are used when $\omega_{\mathrm{p}}/t_h=0.5$ while $t_0=6T_{\mathrm{p}}$, $\sigma_{\mathrm{p}}=T_{\mathrm{p}}$, and $\sigma_{\rm w}'=0.08T_{\mathrm{p}}$ are used when $\omega_{\mathrm{p}}/t_h=0.25$.
The horizontal red and orange lines indicate the exciton energy and Mott gap, respectively.
}
\label{fig:subcycle}
\end{figure}

In Fig.~\ref{fig:HHG_iTEBD}, the exciton energy is close to the Mott gap and the contributions from the exciton are not well distinguished.
Hence, to see excitonic effects clearly, we address the ${\rm large}\mathchar`-U$ regime, in which we can separate the exciton peak from the Mott gap with a strong $V$~\cite{PhysRevB.67.075106}.
Because HHG responses are relatively weak in the large Mott gap system, we employ the ED method for the accuracy of the numerics.
Figure~\ref{fig:HHG_ED} is the calculated HHG spectrum $I(\omega)$ for $U/t_h=24$ and $V/t_h=8$.
Here, the exciton energy (red line) is evaluated by the linear optical response function in the ten-site system and the Mott gap (orange line) is $\Delta_{\rm M}(L=10)$.
These two energies are well separated at $U/t_h=24$ and $V/t_h=8$.
In Fig.~\ref{fig:HHG_ED}, the lower edge of the intensity region at the odd harmonics gradually develops with $\omega$ below the gaps.
The HHG response is strongly enhanced around the exciton energy at $E_0/t_h \gtrsim 1.2$ and the plateau region emerges above the Mott gap.
The intensity region near the exciton energy broadens with increasing $E_0$.
While the exciton energy is well separated from the Mott gap, we find the significant enhancement of the HHG response around the exciton energy, indicating that the exciton in the MI favorably contribute to HHG.

In order to obtain insights into the dynamics of the exciton in HHG, we carry out a subcycle analysis, where we perform a windowed Fourier transformation~\cite{PhysRevB.103.035110} $J(\omega,t)=\int_{t - T_{\mathrm{p}}/2}^{t + T_{\mathrm{p}}/2}J(t')F_{\rm window}(t'-t)e^{i\omega t'}dt'$ using a short window function $F_{\rm window}(t'-t)= \frac{1}{\sqrt{2\pi}\sigma_{\rm w}'} \exp\left[-\frac{(t'-t)^{2}}{2\sigma_{\rm w}'^2}\right]$ and evaluate a time-resolved spectrum $I(\omega,t)=|\omega J(\omega,t)|^2$.
We set $\sigma_{\rm w}'$ in the window function to be much smaller than $T_{\rm p}$ used in the integral range.
$\sigma_{\rm w}'$ in $F_{\rm window}(t'-t)$ may be associated with an experimental probe resolution.
Figure~\ref{fig:subcycle} shows the calculated $I(\omega,t)$ at two different frequencies $\omega_{\rm p}/t_h=0.5$ and 0.25
Corresponding to $I(\omega)$ in Fig.~\ref{fig:HHG_ED}, we find the noticeable response in the sub-Mott-gap regime.
In particular, the time-resolved spectrum $I(\omega,t)$ around the exciton energy (red line) oscillates associating with the driving electric field $E(t)$.
This behavior is clear when $\omega_{\rm p}$ is small.
In Fig.~\ref{fig:subcycle}(b), the intensity region in $I(\omega,t)$ splits into two levels when $E(t)\ne 0$ and they oscillate around the exciton energy, where the energy splitting is maximized at the crest of $|E(t)|$.

\section{Discussion}
The subcycle feature in Fig.~\ref{fig:subcycle} emerging around the exciton energy is qualitatively different from the dynamics of the free doublon and holon at $V=0$.
In the case of the MI without $V$, the motion of the excited free doublon gives rise to the oscillation of the intensity region of $I(\omega,t)$ as reported in Ref.~\cite{PhysRevB.103.035110}.
In this dynamics, the amplitude of the free doublon motion is maximized at $E(t)=0$, i.e., when $|A(t)|$ is maximum. This is because the semiclassical picture based on the dispersion relation of the single-particle spectrum is valid as in conventional semiconductors even though the quasiparticles are replaced by the doublon and holon~\cite{PhysRevB.103.035110}.
However, the subcycle feature around the exciton energy shown in Fig.~\ref{fig:subcycle} does not exhibit the same behavior as the simple semiclassical picture.
For example, in Fig.~\ref{fig:subcycle}(b), the splitting of the intensity region of $I(\omega,t)$ is minimized at $E(t)\sim 0$.
In the case of the MI with $V$, the doublon and holon form the bound state by $V$ and the relative motion of the doublon and holon is strongly restricted.
Hence, the kinematic trajectory of the excited doublon/holon may not be an essential cause of the dynamical feature in Fig.~\ref{fig:subcycle}.
On the other hand, the subcycle feature in Fig.~\ref{fig:subcycle} may also not be caused by the motion of the exciton because the total momentum of the doublon and holon (i.e., the motion of the center of the exciton) should be zero and conserved in the optical excitation in the long-wavelength limit.

The subcycle feature in Fig.~\ref{fig:subcycle} is most likely related to a Stark effect of an exciton.
The extended Hubbard model possesses the odd- and even-parity exciton states in its excited states~\cite{PhysRevB.56.15025}, where we denote them $\ket{\psi_{\rm ex}^{\rm (o)}}$ and $\ket{\psi_{\rm ex}^{\rm (e)}} $, respectively.
While we only observe the odd-parity exciton in the linear response, we may find the contribution from the even- parity exciton in higher-order electric responses.
When the exciton energies are well separated from the others, because of the matrix element $\braket{\psi_{\rm ex}^{\rm (e)} | \hat{x} | \psi_{\rm ex}^{\rm (o)} }$, a strong electric field $E$ may give rise to the hybridization of these two exciton states as $\ket{\psi_{\pm}} \sim c^{\rm(o)}_{\pm} \ket{\psi_{\rm ex}^{\rm (o)}}+c^{\rm(e)}_{\pm} \ket{\psi_{\rm ex}^{\rm (e)}}$, whose energy levels are split by $E$, as in the Stark effect~\cite{Sakurai:1167961}.
While the above discussion is precise in the static limit, this idea may give an interpretation for the instantaneous feature of Fig.~\ref{fig:subcycle}(b), in which the dynamics is relatively slow.
When $E\ne 0$, the both $\ket{\psi_{+}}$ and $\ket{\psi_{-}}$ contain the odd-parity component due to $c_{\pm}^{\rm (o)}\ne 0$ so that the transitions between the ground (even-parity) state and two split states $\ket{\psi_{+}}$ and $\ket{\psi_{-}}$ are optically allowed.
This may correspond to the emergence of two split peaks at $E(t)\ne0$ in Fig.~\ref{fig:subcycle}(b).
On the other hand, when $E=0$, one of two states becomes the purely even-parity state, which is optically inactive.
Actually, in Fig.~\ref{fig:subcycle}(b), single peak only appears at the exciton energy when $E(t)=0$.
Hence, we may interpret the energy-level splitting around the exciton energy observed in Fig.~\ref{fig:subcycle} as the Stark effect of the exciton.

\section{Summary}
We have investigated the excitonic effect on HHG in the one-dimensional extended Hubbard model.
We have found that the HHG spectrum $I(\omega)$ exhibits not only the plateau structure above the Mott gap but also the enhancement of the intensity near the sub-Mott-gap exciton energy.
Moreover, our subcycle analysis shows that the noticeable intensity region in $I(\omega,t)$ around the exciton energy splits into two levels following the driving electric field.
This exciton dynamics is qualitatively different from the dynamics of free doublon and holon since they are bound by interaction $V$.
We suggest that the splitting of the intensity region can be interpreted by the Stark effect of the exciton.
Our calculations have demonstrated that the exciton in the MI favorably contributes to HHG.
\\

\section*{Acknowledgments}
The authors would like to thank S. Ejima, A. Koga, Y. Murakami, and M. Sato for fruitful discussions.The iTEBD and the density-matrix renormalization-group calculations were
performed using the ITensor library~\cite{itensor}.
This work was supported by Grants-in-Aid for Scientific Research from JSPS (Grants No. JP17K05530, No. JP18K13509, No. JP19K14644, No. JP20H01849, and No. JP21K03439). M. U. acknowledges financial support from QS-Fellowship of Chiba University.

\bibliography{References}

\begin{thebibliography}{67}
\bibitem [{cha()}]{cha}%
  \BibitemOpen
  \href@noop {} {}\bibinfo {note} {The light velocity $c$, the plank constant
  $\hbar$, and the lattice constant are set to 1.}\BibitemShut {Stop}%
\bibitem [{dam()}]{damp}%
  \BibitemOpen
  \href@noop {} {}\bibinfo {note} { In practice, the damping factor
  may be regarded as the dissipation or scattering by other parts of real
  solids, e.g., phonons and impurities, which are not incorporated in our
  model.}\BibitemShut {Stop}%
\end{thebibliography}

\end{document}